\pdfoutput=1

\documentclass[utf8]{arxiv}

\usepackage{url,hyperref,lineno,microtype}
\usepackage[onehalfspacing]{setspace}

\usepackage{soul}
\usepackage{booktabs}
\usepackage{multirow}
\usepackage{pifont}
\usepackage{colortbl}
\usepackage{subcaption}
\usepackage{graphicx}
\usepackage{caption}
\usepackage{wasysym}
\usepackage{cleveref}
\usepackage{tabularx}

\newcommand{\eg}{e.\,g., }
\newcommand{\ie}{i.\,e., }

\usepackage[acronym, shortcuts, nohypertypes={acronym}]{glossaries}

\newacronym{AI}{AI}{artificial intelligence}
\newacronym{CNN}{CNN}{convolutional neural network}
\newacronym{DL}{DL}{deep learning}
\newacronym{FER}{FER}{facial expression recognition}
\newacronym{LSTM}{LSTM}{long short-term memory}
\newacronym{MFCC}{MFCC}{Mel-frequency cepstral coefficients}
\newacronym{NLP}{NLP}{natural language processing}
\newacronym{RF}{RF}{random forest}
\newacronym{RNN}{RNN}{recurrent neural network}
\newacronym{SVM}{SVM}{support vector machine}

\def\keyFont{\fontsize{8}{11}\helveticabold }
\def\firstAuthorLast{Schuller {et~al.}} 
\def\Authors{Bj{\"o}rn W.\ Schuller\,$^{1,2}$, Shahin Amiriparian\,$^1$, Anton Batliner\,$^1$, \\Alexander Gebhard\,$^1$, Maurice Gerzcuk\,$^1$, Vincent Karas\,$^1$, Alexander Kathan\,$^1$, Lennart Seizer\,$^3$, Johanna L\"ochner\,$^3$}
\def\Address{$^{1}$EIHW -- Chair of Embedded Intelligence for Health Care and Wellbeing, University of Augsburg, Augsburg, Germany \\
$^{2}$GLAM -- Group on Language, Audio, \& Music, Imperial College London, London, United Kingdom \\
$^{3}$Department of Child and Adolescent Psychiatry, Eberhard-Karls University, T\"ubingen, Germany}
\def\corrAuthor{Bj\"orn W.\ Schuller, Johanna L\"ochner}

\def\corrEmail{schuller@ieee.org, Johanna.Loechner@med.uni-tuebingen.de}

\begin{document}
\onecolumn
\firstpage{1}

\title[Computational Charisma]{Computational Charisma -- \\A Brick by Brick Blueprint for Building Charismatic Artificial Intelligence}

\author[\firstAuthorLast ]{\Authors} 
\address{} 
\correspondance{} 

\extraAuth{}

\maketitle

\begin{abstract}

\section{}
Charisma is considered as one's ability to attract and potentially also influence others. Clearly, there can be considerable interest from an artificial intelligence's (AI) perspective to provide it with such skill. Beyond, a plethora of use cases opens up for computational measurement of human charisma, such as for tutoring humans in the acquisition of charisma, mediating human-to-human conversation, or identifying charismatic individuals in big social data.
While charisma is a subject of research in its own right, a number of models exist that base it on various `pillars', that is, dimensions, often following the idea that charisma is given if someone could and would help others. Examples of such pillars, therefore, include influence (could help) and affability (would help) in scientific studies or power (could help), presence, and warmth (both would help) as a popular concept. Modelling high levels in these dimensions, i.\,e., high influence and high affability or high power, presence, and warmth for charismatic AI of the future, e.\,g., for humanoid robots or virtual agents, seems accomplishable. 
Beyond, also automatic measurement appears quite feasible with the recent advances in the related fields of Affective Computing and Social Signal Processing. 
Here, we, thereforem present a blueprint for building machines that can appear charismatic, but also analyse the charisma of others. 
To this end, we first provide the psychological perspective including different models of charisma and behavioural cues of it. We then switch to conversational charisma in spoken language as an exemplary modality that is essential for human-human and human-computer conversations.
The computational perspective then deals with the recognition and generation of charismatic behaviour by AI. 
This includes an overview of the state of play in the field and the aforementioned blueprint. We then name exemplary use cases of computational charismatic skills before switching to ethical aspects and concluding this overview and perspective on building charisma-enabled AI -- will tomorrow's influencers be artificial?


\tiny
 \keyFont{ \section{Keywords:} Charisma, AI, Empathy, Mimicry, Affective Computing, Social Signal Processing} 
\end{abstract}

\section{Introduction}
\label{sec:introduction}

Charisma. An irresistible force that, apart from beauty or rhetoric, captivates people. A miracle cure for professional success and an almost effortless rise to the top of power. A plethora of popular science literature, podcasts, and discussions rotate around this fascination, providing training to adopt a charismatic style -- going along with the great promise of being successful and attractive to others. Besides this great uptake, the topic of charismatic behaviour also has a research tradition in sociology and psychology and is now increasingly trending in computation. This is promising, since computational charisma may be applied to numerous fields such as  leadership training, mental health care, and education and enhance outcomes in several ways: more efficient leadership, increased comfort in recipients, better teamwork, and reduction of reluctance and irritation. However, charismatic behaviour is not bound to particular values and initially exists independently of an ideology. This also makes the appropriation of charisma a potential danger if it is misused for unethical purposes. 
In this article -- section by section --, we initially discuss the myth about charisma from a scientific, but also popular science perspective (functional aspects of charisma), to follow up with several layers of markers for charisma (formal aspects of charisma), computational aspects of charisma to finish with the motivation based on applications 
and rendering these more attractive and effective 
and ethics of computational charisma.
Thereby, we focus on spoken language and audio as indelible key features of charismatic behaviour, that play undoubtedly a key role in times of remote and digital -- hence restricted visual -- communication. 





\section{Functional Aspects of Charisma: Psychological Models}

Although charisma is a ubiquitous and frequently discussed phenomenon, and people seem to agree on which person inherits this trait, a certain mysteriousness surrounds the exact definition. And yet, although some popular figures in history are frequently characterised as being charismatic. In this first section, we discuss charisma from a sociological, psychological and also popular science perspective and aim to untidy the concrete characterisations of charisma. In contrast to other research fields, the subjective perception of charisma and popular science uptake of this phenomenon is particularly interesting, since it is part of its definition and hence, immanent.

\label{sec:psychos}


\begin{figure*}[tb!]
	\begin{center}
		\includegraphics[width=\textwidth]{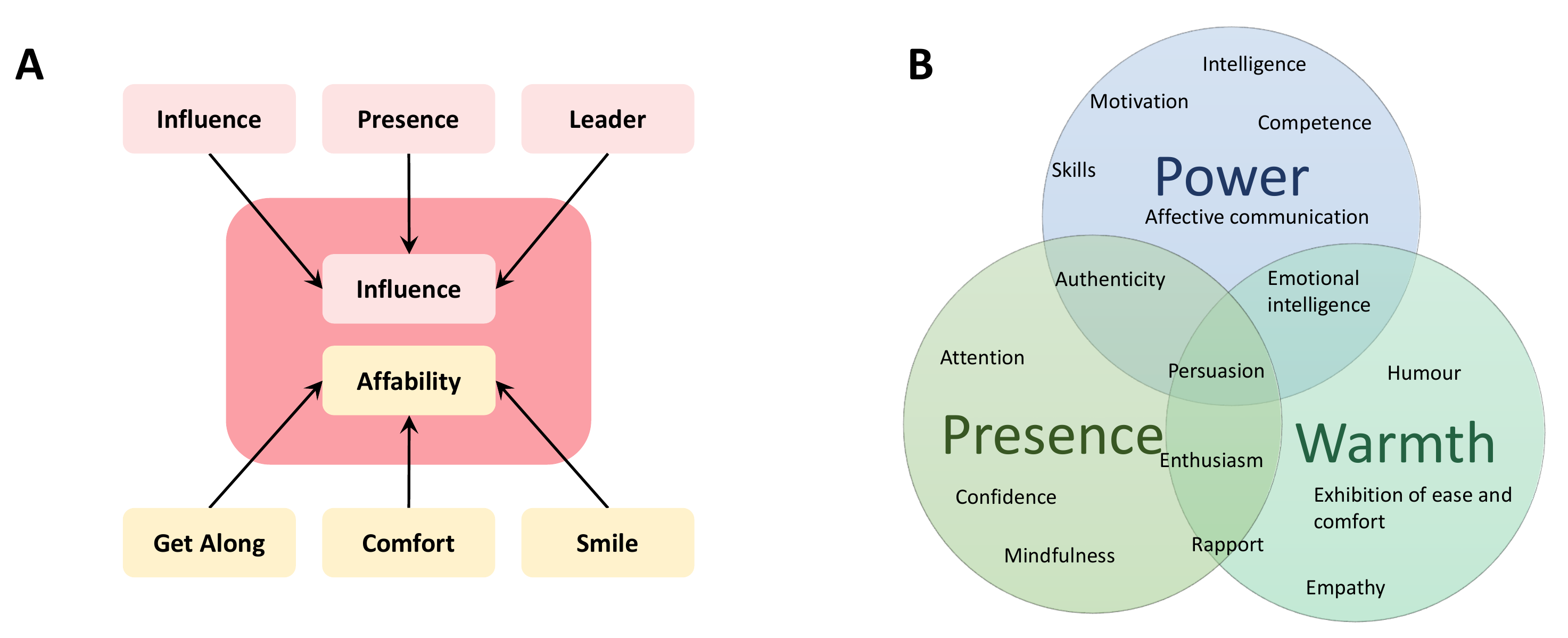}	
		\caption{Comparison of the models of Tskhay (A)~\citet{Tskhay18-CIE} and Fox Cabane (B)~\citet{Foxcabane13_TCM}}
		\label{fig:tskhay_and_cabane}
	\end{center}
\end{figure*}
    

\subsection{Origin and Definition}


The word charisma originally comes from the Greek 
($\chi\acute{\alpha}\rho\iota\sigma\mu\alpha$)
and means `gift of grace'. Even the ancient Greeks assumed that charisma is a gift from God that some have, and others do not. Today, the word is typically used as a descriptor for people who are attractive to others and manage to gather a following around them (for the good or bad) such as Princess Diana, Oprah Winfrey, Martin Luther King, or Adolf Hitler. Although everybody has an intuitive understanding of the concept of charisma and there is a high agreement in the population about which persons are charismatic, a scientifically sound and commonly used definition is still discussed \citep{Antonakis12-TIC}. This may partly be explained by the fact that the study of charisma is relatively young and still mostly restricted to economic psychology in terms of leadership research. Naturally, children are even less likely to have a defined concept of what charisma actually means; however, they are well capable of voting the `captain'. Hence, Antonakis and Dalgas asked 5-13 years-olds to rate the `captain' among a selection of pairwise displayed photographs of French candidates for the presidency, resulting in an 85\,\% hit rate \citep{Antonakis09_PEC}. Although these results rely on visual data only, it was shown that attractiveness is not the key feature of charisma. For example, in a study predicting success in relationships and academic careers, charisma was more predictive than attractiveness and visuals \citep{Orzeatja11_LAT}. 
The sociologist Max Weber defined charisma at the beginning of the 20th century  as an `extraordinary quality of a personality', as a `supernatural or superhuman power' \citep{Weber22-GDS}. Based on this work, House \citep{House96-A1T} provided the first operationalisation. Thereby, charisma was defined as the ability to inspire others toward a common goal and identity by appealing to their emotions and collective identity in order to impart an idealised vision to their followers -- thus, the central role of charisma in leadership research. In the following decades, more specific traits and behaviours have been associated with charisma \citep{Antonakis12-TIC}. A key quality lies in the ability to connect with other people and exhibit ease, trust, and comfort in the audience paving the ground to become a leader. In addition, a charismatic person is highly persuasive \citep{Mhatre14_CAT}. Several definitions of such phenomena are discussed in the literature including properties such as authenticity, emotional competence (e.\,g., understanding emotions in oneself and others, or managing own emotions), empathy, persuasiveness, spending attention to others, passion, enthusiasm, and obtaining strong opinions to name a few. 
However, such qualities may not only be used for charity aspects. Welsh and colleagues investigated the associations of psychopathy, charisma and success. They found that psychopathy was positively associated with leadership charisma and the influence component of general charisma \citep{Welsh2021_PCA}; in addition, charisma moderated the association of psychopathic traits and perceived success in form of the evading detection and punishment.

Charisma models or concepts were proposed as either inherent personality traits \citep{Burke1981_CCN}, observer perception and outcomes \citep{Awamleh99_POL}, or both \citep{Conger09_MCD}.
Aiming to provide a more comprehensive model of charisma based on empirical data, \citet{Tskhay18-CIE} created an empirical model of charisma: They investigated characteristics of charisma by rigorous and repeated questioning of people how they use to describe charismatic people, and subsequently applied factor analyses to identify the most important components. Their analyses resulted in a two-factor model with one factor -- \textit{influence} -- consisting of items that describe leadership ability and influence in a group setting, and another factor -- \textit{affability} -- that consists of items describing a pleasant and inviting disposition toward others. The factors with more detailed descriptions and an exemplary list of behaviours that are associated with each are given in 
this section, 
with no claim to completeness. While \textit{influence} and \textit{affability} are separate qualities, somehow, in the combination of traits and behaviours associated with these two, charisma emerges as a novel trait. Charisma is thereby defined as a multi-dimensional construct of traits and behaviours in contrast to `just' being a likeable person. Similarly, \citet{Keating11-CCT} argued that dominant behaviour triggers avoidance reactions in others, whereas emotionally warm behaviour triggers approach reactions.
She further claims that the perception of charisma emerges specifically through the simultaneous elicitation of avoidance and approach reactions by the combination of \textit{influence}(dominance, power) and \textit{affability} (emotionality, approachability) in a charismatic person.


\begin{table}[ht]
\centering
\begin{tabularx}{\linewidth}{p{7mm} p{20mm} p{57mm} p{78mm}}
     \multicolumn{4}{c}{} \\
     \toprule
     & Model 
     & GCI Item
     & Associated constructs
     \\
     \midrule
     \rotatebox[origin=r]{90}{Influence}
     & 
     Influence \newline
     \newline Presence \newline
     \newline Leader
     & 
     Has the ability to influence people \newline
     \newline Has a presence in a room \newline
     \newline Knows how to lead a group
     & 
     \textbf{Convergent:} Emotional Intelligence, Positive Affect, Extraversion, Openness, Conscientiousness, Political Skill, 
     \newline Competence
     \newline \textbf{Discriminant:} Negative Affect, Neuroticism 
     \\
     \midrule
     \rotatebox[origin=r]{90}{Affability}
     & 
     Get along \newline
     \newline Comfort \newline
     \newline Smile
     & 
     Can get along with everyone \newline
     \newline Makes people feel comfortable \newline
     \newline Smiles at people often
     & 
    \textbf{Convergent}: Emotional Intelligence, Positive Affect, Confidence, Extraversion, Openness, Conscientiousness, Agreeableness, Political Skill, Competence, Warmth
    \newline \textbf{Discriminant}: Negative Affect, Neuroticism
     \\
     \bottomrule
\end{tabularx}
\label{tab:tskhay_model}
\caption{Detailed overview of the Tskhay et al.\ model}
\end{table}

There are several psychological constructs that may be convergent or discriminant to charisma. In a validation study of the influence-affability model, the uniqueness or relatedness of charisma to other individual difference measures was tested in multiple samples (\citep{Tskhay18-CIE}; see Table 1). Thereby, \textit{influence} and \textit{affability} were both found to be significantly related to emotional intelligence, \ie the appraisal, expression, regulation, and utilisation of emotions in a variety of contexts \citep{Schutte98-DAV}. In terms of emotionality experienced by oneself, positive affect was positively related to the two charisma factors, while negative affect was negatively related to both. Political skill, defined by the four dimensions of social astuteness, interpersonal influence, networking ability, and apparent sincerity \citep{Ferris05-DAV}, is often used as a metric of charismatic leadership and accordingly was found to be positively related to \textit{influence} and \textit{affability}. Intelligence is a trait that is often ascribed to charismatic individuals in lay theories; however, intelligence, as determined by Raven's Matrices \citep{Raven98-RPM}, was not associated with influence or affability, indicating that charisma may rely more on interpersonal skills in social interactions rather than intelligence. Further, the general confidence of an individual as the degree to which one feels certain about both the world and idiosyncratic surroundings and their ability to deal with stress \citep{Keller11-TGC} was positively related to affability, but not associated with influence. In terms of personality traits  as the Big Five  \citep{Mccrae99-HOP}, openness, consciousness, and extraversion were positively related to both influence and affability, while agreeableness was only related to affability. Neuroticism on the other hand was negatively associated with influence and affability. Of the dimensions competence and warmth -- two essential elements of both social behaviour and personal characteristics \citep{Fiske07-UDO} -- influence was only related to competence, while affability was related to both warmth and competence.

Another model of charisma was proposed by \citet{Foxcabane13_TCM}, in which she refers to charisma as deriving from three pillars: \textit{presence},  \textit{power}, and \textit{warmth}. \textit{Presence} is displayed by dwelling in the current moment, active listening, and responding adequately. The focus of attention lies on the person one is talking to and taking an honest interest in the conversation partner. \textit{Power} is not defined as actual power like being in a position as president or high-rank manager. It is rather understood as high competence due to certain skills, abilities, or intelligence a person obtains. \textit{Warmth} requires a high level of empathy, openness, and positivity. The pillar \textit{warmth}  has frequently been studied as part of the two-dimensional \textit{warmth} and \textit{competence} (W\&C) model \citep{Wang2021_AOD, Fraser2021_UAC, Fraser2022_CMO}, where \textit{warmth} indicates the nature of the sender's intent towards the receiver, and \textit{competence} the ability of the sender to enact this intent. The combination of these dimensions evokes emotional responses ranging from admiration to disgust \citep{Fraser2022_CMO}. 
Thus, \textit{warmth} is closely related to the perceptions of attractiveness and empathy. Therefore, charismatic individuals usually radiate acceptance and friendliness that one otherwise experiences only from family members or friends. It is discussed whether one or two of the three qualities may be sufficient to appear charismatic, as Steve Jobs, for instance, scored with \textit{presence}  and \textit{power}, but lacked \textit{warmth} . 
In contrast, Martin Luther King showed all three qualities. Hence, the pillars \textit{warmth} and \textit{power}  may relate to \textit{affability}  and to the \textit{influence} of the two-factor model by Tskhay et al.\ \citep{Tskhay18-CIE}, while the pillar \textit{presence}  was discussed as a non-latent, \ie secondary, variable in their empirically found model by a factor analysis following questioning participants (see Figure 1A). 
  
Very similarly, the concept of \textit{rapport}  is defined and may well serve as part of charismatic behaviour. \citet{Tickle-Degnen90_TNO} conceptualised the nature of \textit{rapport}  in terms of a dynamic structure of three interrelating components: mutual attentiveness, positivity, and coordination; these are differently weighted and present over time in a relationship. Hence, \textit{rapport}  is characterised by agreement, mutual understanding, or empathy that makes communication possible or easy, establishing ease and comfort in communication partners. In consequence, a charismatic individual is capable of establishing rapport.

Conclusively, charisma is a person-specific descriptor that emerges specifically in social situations through the attribution of a certain set of traits to an individual. Despite heterogeneous conceptualisation and the inherent complexity, there is a consensus that charismatic individuals exert influence over others, have extraordinary social skills, comfort and connect to others,  inspire followership,  and are prone to leadership roles \citep{Tskhay18-CIE}\citep{Antonakis12-TIC}. 
Breaking charisma down to such concrete properties reveals a combination of personality traits and skills that are partly inherited, socially acquired, and trained. In social psychology, the processes that leads us to form impressions about other people are referred to as person perception~\citep{Moskowitz13-PP}. Some methods of perceiving another person involve inferring details about them based on observations of their activities. Other types of personal perception happen more immediately and only need one to view another person. In building a machine that people perceive as charismatic, a bias in human inference processes can be exploited, namely the fundamental attribution error: People tend to ascribe observed behaviours to internal factors like personality or character rather than to external factors such as situational constraints \citep{Colman16-GTA}. Thus, by mimicking certain appearance cues, characteristics, and behaviours programmatically to elicit the perception of charisma-associated traits, it should be possible to build a ``charismatic artificial intelligence''.


\subsection{Acquisition of Charismatic Behaviour}

In consequence, charismatic behaviour can be acquired and there is a plethora of trainings offered especially in the field of leadership coaching. 
Overall, the two key qualities, \ie factors, introduced by \citet{Tskhay18-CIE} may be achieved especially through confidence and skills (\textit{influence} ), emotional intelligence, and empathy (\textit{affability} ).
 As is conclusive from the above (see also Figure 1), they may also be complemented 
 by a third pillar or factor suggested by \citet{Foxcabane13_TCM} -- mindfulness (\textit{presence}). Following  \citet{Foxcabane13_TCM} and  \citet{Tskhay18-CIE}, such characteristics will elicit an increased impression of attractiveness, energy, persuasiveness, power, and empathy, and establish rapport between communication partners. 

Focusing on leadership trainings, and  translating charismatic behaviour into more concrete tactics, \citet{Antonakis12_CCB} investigated  twelve techniques to increase charisma -- the so-called ``charismatic leadership tactics'' (CLTs). Similar to athletes who follow a training schedule, leaders who aim to become influential, trustworthy and ``leaderlike'' are recommended to practice certain tactics regularly.
For this purpose, they examined the nomination speeches of all candidates for president in the U.\,S.\ between 1916 and 2008. The analysis revealed that the use of figurative language, anecdotes, proverbs, and the proper use of body language had a significant impact on the outcome of the election. Despite humour, repetition, and talking about sacrifices, such verbal and non-verbal techniques were shown to have the greatest impact 
in any context. The nine verbal techniques are metaphors, similes, and analogies; stories and anecdotes; contrasts; rhetorical questions; expressions of moral conviction; reflections of the group's sentiments; three-part lists; the setting of high goals; and conveying confidence that they can be achieved. 
For example, the metaphor of being on a boat in a storm may serve as a metaphor for a critical period. Even without being a born raconteur one can tell the compelling story of taking a deep breath as ``anchor'' and visualise the north star for guidance. Another example to motivate followers through a challenging period would be an anecdote of a personal story, as climbing a mountain when a thunderstorm arises and how the team must have kept going. In addition, there are three non-verbal techniques: animated voice, facial expressions, and gestures. Keeping with voice-associated techniques to improve oneself's charisma, or rather the perception of charisma in others, 
it is suggested to speak clearly, fluently, forcefully, and in an engaging manner that invokes images, energy, and action; moreover, the delivery's pace and intonation should be varied, with a general upbeat tempo and an occasionally slowing down to create tension \citep{Tubbs19-LC}. Similarly, \citet{Foxcabane13_TCM} recommends lowering the tone of one's voice at the end of each statement and make frequent pauses while speaking. Despite these strategies to develop or improve charisma, the debate on whether charisma can be learnt or simply is a trait with set between-subject variation is still ongoing \citep{Tubbs19-LC}. 
It is of note that, even in human generation of charisma, 
an attribution error can apply: when speakers learn to speak with a `charismatic voice', people perceive them as charismatic, even when their personality does not change. 

Hence, very concrete acquired behaviour was shown to lead to a more charismatic behaviour of individuals and hence, can be installed on who- or whatever to some extent: A person  might not be charismatic in themselves  but may appear this way, due to, \eg speaking in a charismatic way. Note that it was shown that appearance is not the key factor in `charismatic appearance'.
 \citet{Antonakis12_CCB} observed that in eight out of ten U.\,S.\ presidential races, candidates who deployed such verbal CLTs won more often. 
Since communication nowadays is primarily technology-mediated, \citet{Ernst22_VCL} investigated CLT in a recent prospective meta-analysis on virtual charismatic leadership.  The meta-analytic effect of CLTs on performance (Cohen's $d$ = 0.52 in-person, $k$ = 4; Cohen's $d$ = 0.21 overall, $k$ = 10) and engagement in an extra-role task (Cohen's $d$ = 0.19 overall; $k$ = 6) indicated large to moderate effects. Yet, for performance in a virtual context, Cohen's $d$ ranged from $-0.25$ to 0.17 (Cohen's $d$ = 0.01 overall; $k$ = 6).
 In summary, disentangling especially phonetic and linguistic markers for charisma may be particularly beneficial in times of virtual communication in all kind of fields.

In the following section, this article focuses on the specific phonetic, linguistic, and other markers that are associated with charisma.



\section{Formal Aspects of Charisma: Phonetic,  Linguistic, and Other Markers}
\label{sec:formal_aspects_of_charisma}

The marking of charisma is definitely multi-modal, and trading relations exist both between and within modalities -- \ie a more pronounced but not exaggerated marking in one parameter can compensate for weak signalling in another parameter, see \citet{Niebuhr20-ACP}. Ranking the importance of modalities is futile and either based on intuition or on one or only a few studies with their specific databases, designs, and methods; see the `7\,\%-myth' \citep{Mehrabian67-DOI} and \cite[p.~14]{Schuller14-CPE}.
 In the following, we concentrate on speech and language, \ie on vocal and verbal parameters. This appears reasonable, given the focus on today's AI often communicating with users by this modality; in addition, also when analysing human interaction, spoken language plays a key role. 
 Yet, we will as well present a sketchy overview of charisma conveyed within the other modalities. We will start with phonetic markers of charismatic speech in~\Cref{Phonetic-Markers}; then follow linguistic markers in~\Cref{Linguistic-Markers}, and other modalities~\Cref{Other-Modalities}.



The intuitive understanding of charisma is mirrored in equally intuitive characterisations such as attractive, inspiring, animated, enthusiastic, warm, likeable, or pleasant. In this section, we now report the state of the art in mapping these terms onto markers that can be measured and counted.

\subsection{Phonetic Markers}
\label{Phonetic-Markers}

Arguably, Rosenberg and Hirschberg described the first sets of studies on charismatic speech  \citep{Rosenberg09-CPF}. So far,  most of them addressed charisma in politics (candidate speeches) and Marketing  \citep{Zoghaib19-POV} and concentrate on prosody.
Related are states and  traits such as
leadership \citep{Weninger12-TVO}, 
competence/trustworthiness \citep{Yang20-WMA,Davidson21-TVO},
likability \citep{Weiss10-VAA,Schuller15-ASO}, and 
(sexual) attractivity \citep{Trouvain20-VAC}.
Charisma
can be tied to performing something, \eg a candidate speech, and can be switched on and off; 
see  \citet{Rosenberg09-CPF}: 
``Speakers were rated as more charismatic when
they were delivering a stump speech (mean rating of 3.28) than when they are being interviewed (2.90).''
So, at least the `acoustics of charisma' are not in an `always one-to-one relationship to personality. Of course, this makes it possible for it  to be taught and acquired. 
An infamous example is Adolf Hitler where the only recording of non-public speech (\url{https://www.youtube.com/watch?v=WE6mnPmztoQ}) reveals a relaxed, almost  likeable  style of speaking, much different from his public speeches. We can distinguish `dark charisma' \citep{Fragouli18-TDS},  where, \eg anger can be strongly marked with prosodic means, when this is in accordance with the audience, from  `bright charisma' which can be rather marked prosodically (Barack Obama) or linguistically and by the context (Mahatma Gandhi); see \cite{Errico13-TPO}. 
Even psychopaths can display traits of bright charisma 
in discordance to their personality \citep{Weatherby16-TMO}.
Intervening factors can be gender, age, and culture  \citep{Errico13-TPO}. Laryngealised, `creaky' voice 
  -- that is at the same time indicating very low but also irregular pitch --  can make men more cool and attractive   \citep{Davidson21-TVO}, whereas a breathy voice is preferred for women
\citep{Greer15-TPO}; this, however, mostly holds for younger women \citep{Anderson14-VFM}, 
whereas in business and academia, a creaky voice can be a sign of competence for both females and men. \citet{Klofstad16-HVP} summarise the experiment on leadership: 
``... males with lower-pitched voices tend to be perceived as more attractive, physically stronger, and more `dominant' 
... 
For females, the standard is dichotomous: Women with higher-pitched voices tend to be considered more attractive, whereas those with lower-pitched voices are perceived as more dominant.''; see as well \cite{Anderson12-PFL,Klofstad15-POC,Zoghaib19-POV}. 
%
%
\citet{Niebuhr2020_WMB} compared customer and investor keynotes of Steve Jobs and Mark Zuckerberg. 
Jobs, commonly perceived as the more charismatic speaker, produced a higher pitch level (even approaching that of female speakers), and almost twice the pitch range of Zuckerberg. Jobs used shorter phrases, had fewer disfluencies, and scored higher in the voice quality metrics. However, he did not exceed Zuckerberg in terms of intensity variability. Both showed significant differences when addressing customers and investors, showing again that charisma is situation-dependent. 

Low-level descriptors of the voice have been shown to convey perceptions of speaker personality traits \citep{Schuller15-ASO}. The likeability of a person can be predicted using pitch frequency F0, articulation rate, and spectral parameters such as MFCC \citep{Weiss10-VAA}. \citet{Errico13-TPO} conducted a cross-cultural study showing the effects of pitch and the duration of speech pauses on the perception of two dimensions aggregated from 67 traits and conforming to proactivity-attractiveness and calm-benevolence.

As far as prosody is concerned, we can sum up with \citet{Yang20-WMA}: ``...  voices that are louder, higher, faster, and with greater ﬂuctuation in pitch were rated as more charismatic.'' 
Now, we `only' have to define the acceptable range of these prosodic varieties; 
too great an intensification will certainly yield undesirable consequences such as the impression of distortion or a lower discriminability, see \citep{Hamilton93-EAI}, \citep{Holz21-TPR}.
Moreover, higher pitch range and 
overall, more variability characterising charismatic speech, differ from less variability and lower pitch, characterising competence; see again  \citep{Rosenberg09-CPF}.
Other prosodic parameters as well, and  other acoustic parameters such as spectral distributions, favourable for conveying charisma, can be described as `well-balanced' and `well-shaped': neither too integrating nor too isolating prosodic phrasing -- \ie not too many but not too few pauses;  more spectral energy at low frequencies (`full voice'); and more precise articulation (no centralisation of vowels).
A charismatic voice is definitely not characterised by dysphonia, \ie disordered voice  (hypophonia, \ie soft voice, or the opposite, hyperphonia, \ie tense, harsh voice).
%
%
Based on all these findings, \citet{Niebuhr20-ACP} describe a system for charisma profiling.

\begin{figure*}[tb!]
	\begin{center}
  		\includegraphics[width=1\textwidth]{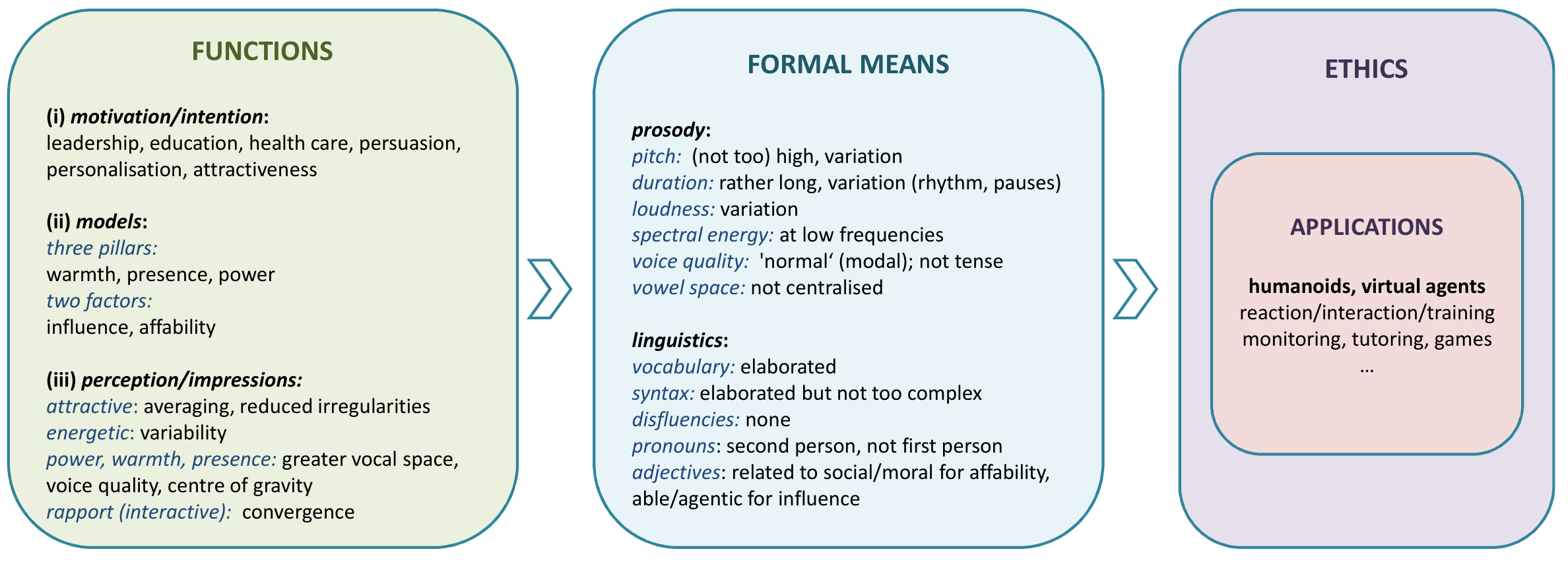}
		\caption{Overview of concepts and components: three stage \textbf{FUNCTIONS} of charisma; (i) higher level \textit{motivation/intentions} employ (ii)  \textit{models} (three \textit{pillars} and two \textit{factors}) to create (iii) specific \textit{perception/impressions} conveyed via speech by using \textit{prosody} and \textit{linguistics} (\textbf{FORMAL MEANS}); this charisma is then used in  \textbf{APPLICATIONS} in human-machine-interactions that \textbf{ETHICS}  has to take care of.}
		\label{fig:charisma-components}
	\end{center}
\end{figure*}

\Cref{fig:charisma-components} summarises the formal acoustic aspects dealt with in this subsection and the linguistic aspects described in \Cref{Linguistic-Markers}; at the same time, it relates these formal aspects to the functional aspects: the motivation behind creating charismatic agents; the models employed by us; and the perception and impression that such charismatic agents have on the human interaction partners. These  components are employed to create applications where charisma is harnessed to achieve their specific goals. Ethics has to assess and possibly restrict the use of charisma in these applications, 
see \Cref{ethics}.

\subsection{Linguistic Markers}
\label{Linguistic-Markers}




As far as linguistic markers are concerned, the use of informal language, high occurrence of pronouns, and avoidance of synonyms can be used to elicit greater warmth, while the opposite holds for formal, complex language. For pronouns, those that involve the audience, \eg \textit{we} and \textit{you}, are useful for creating a better first impression \citep{Biancardi2019_ACM}. In addition, \citet{Rosenberg09-CPF} found that using first-person pronouns positively correlated with the charisma ratings of political candidates in spoken but not in written form.

Adjectives can serve as markers for the charismatic content of language. They can be clustered via concepts such as sociability and morality for warmth or ability and agency for competence \citep{Fraser2021_UAC}. The usage of adjectives, as opposed to nouns in describing persons, affects the generated impressions, with nouns conveying a greater sense of defining, immutable traits \citep{Fraser2022_CMO}. When referring to groups of people, the choice of descriptor can evoke various impressions of warmth and competence via associated stereotypes; consider, \eg the differences between \textit{the elderly}, \textit{old people}, \textit{old folks} and \textit{senior citizens} \citep{Fraser2022_CMO}.

The clarity of the intended message also affects the perception of charisma. A lower amount of disfluencies may make a speaker appear more confident and focused. The negative effect of disfluency is more pronounced for linguistics than for prosody according to a comparison between speech and transcripts by \citet{Rosenberg09-CPF}, possibly because the audience may expect it in spoken but not in written form. Regarding the content of a message, conveying more information is not necessarily beneficial from a charisma perspective;  
 for 
 speakers with a lower ratio of function to content, words can be rated as more charismatic, possibly due to higher rhetorical complexity \citep{Rosenberg09-CPF}.

Charisma is closely related to being able to influence others, thus, here we also examine linguistic markers of persuasion. \citet{Guerini03-PMF} propose a taxonomy resting on four pillars: cognitive state, social relations, emotional state, and interaction context. Here, cognitive elements refer to goals and beliefs of agents and concepts related to them, social elements deal with power dynamics between relevant persons, emotional elements can be used to enhance or diminish a message, and contextual elements can add useful information. Persuasion strategies are then grouped  by their objective: inducing a change in beliefs, and inducing a change in actions. The former can be achieved by appealing, \eg  to the opinions of experts, to public opinion, or to empirical evidence. The latter may follow a social strategy by appealing to someone from whom the target derives standards or morals, or to the target's social image at large. Another option would be to present imaginary consequences, either positive via promises or negative via threats. A charismatic agent may select and modify these strategies to improve the success rate.

\subsection{Vision, Touch, and More: Markers in Other Modalities}
\label{Other-Modalities}


Charisma without spoken or written language may hardly exist, but obviously, other channels contribute, \eg in the visual modality gestures, body pose, facial expressions, and gaze behaviour. 
While today's interaction and communication with AI is largely focused on spoken and written language, future AI is expected to be doing so multimodally, detecting the user state and responding in real-time to generate a favourable, human-like impression \cite{Biancardi2017a_TAC}.

\citet{Cuddy2008_WAC} investigated how warmth and competence are perceived based on behaviour at interpersonal and intergroup levels. Smiling, as well as engaging gestures, touch, and mirroring were found to increase the impression of presence and warmth, while disengagement and creating physical distance by leaning back or turning away decrease it. Expansive and open body poses, suggesting power and dominance,  resulted in higher impressions of competence. For hand gestures, the use of object adaptors and ideationals (relating to spoken words) improved the speaker competence, while self-adaptors decrease it \citep{Biancardi2017_AFI}.

In general, a speaker's delivery can have a great influence on their credibility, \ie a strong delivery is more likely to lead to high credibility than is a weak one \citep{Holladay1993-CVA}. Factors contributing to a good delivery include eye contact, gestures, and facial expressions \citep{Holladay1993-CVA}. This is not surprising as gestures and facial expressions can innately radiate charisma~\citep{Towler2003-EOF}.
Since these characteristics are settled in the visual domain, they have to be considered apart from audio. 

Regarding conversational interaction, when a person's gaze is focused on the conversation partner, this is a sign of attention and shows both interest in the conversation and commitment to the conversation partner \citep{Knight2013-EHI,Freeth2019-SPG}. That is, if the gaze is wandering through the surroundings it may evoke the impression that a person is not fully listening and wants to distract themselves with seemingly more interesting things.
Thus, to recognise the attentiveness and presence in a conversation, one of the easiest approaches might be to track eye contact and face gaze in general. 

Another tool of nonverbal behaviour and conveying (intimate) emotions is the sense of touch. Touch is crucial for social development and necessary for children in order to grow up healthy \citep{Van2015-STI,Weiss2000-TTC}. Out of all nonverbal modalities, affective touch is our primary channel for expressing intimate emotions and can effortlessly establish social presence \citep{Van2015-STI}. In addition to distinct emotions like love, anger, and fear, touch can also convey more complex emotional patterns such as trust, receptivity, and affection \citep{Van2015-STI,Hertenstein2006-TCD,Hertenstein2009-TCO,Burgoon1991-RMI}. As previously mentioned, charismatic persons radiate characteristics like trustability, presence, and warmth, which makes affective touch an essential modality next to audio
-- if appropriate in the specific situation.

\section{Computational Aspects of Charisma: Modelling}
\label{sec:nerds}
After analysing the markers of charisma in \Cref{sec:formal_aspects_of_charisma}, we now deal in this section  with the modelling of charisma from a computational perspective. Automatic recognition of charisma describes the detection of the sociological and psychological markers for charismatic behaviour using machine learning approaches. Similarly, the automatic generation of charisma outlines methods for generating auditory or visual charismatic traits.

\subsection{Automatic Recognition of Charisma}




Charisma can be registered via a wide range of modalities, ranging from facial movements and gestures to speech and physiological attributes like heart rate and skin conductance. Since charisma is an interpersonal effect, computational analysis can focus either on the sender projecting charisma, on a receiver forming an impression, or on dyadic interactions between the two~\citep{Wang2021_AOD}. Here, we take up the stated sociological, psychological, and popular science perspective and translate it into computational aspects of phonetics, linguistics, and other modalities in automatic recognition of charisma.

\subsubsection{Audio}

The quality of speech transmission has an impact on the perception of charisma. \citet{Gallardo2018} investigated the effect of bandwidth on perception of male and female speakers selected for extreme values of warmth-attractiveness (WAAT). Shifts in traits such as maturity, sympathy, and confidence for males and competence for females can be explained with alterations of F0 resulting from the narrow-band transmission. Another study \citep{Gallardo2019} assessed the impact of various encoding and transmission properties on the binary classification of warmth and attractiveness via Random Forest and Support Vector Machines. Narrowband codecs were found to degrade performance to near chance level. Packet loss also confused the classifiers, while jitter had minor effects.

In the early years of prosody research in automatic speech processing
\citep{Batliner20-PIA}, the focus was on detecting and classifying linguistic phenomena such as phrase accents, boundaries, disfluencies, sentence modality, and dialogue acts; such explicit modelling was then superseded by  implicit modelling in AI approaches. Yet, it might gain momentum in our context, when we want to model markers for charisma.  
Asking questions during a conversation indicates that a person is listening and interested in what the conversation partner says;  thus, it can indicate the \textit{presence} in a conversation. 
In this context, acoustic and phonetic features are deployed, at which lexical features can also be crucial for the correct identification of declarative questions \citep{Ando2018-AQD}. Furthermore, \acp{RNN} are applied in order to obtain the high-level contextual information over time \citep{Tang2016-QDF}.

Before asking a question, it can also be beneficial to make a short pause, in order to show that one thinks about what the conversation partner has said, before giving an answer. This can convince the other person that one is listening carefully and is present in the conversation. \citet{Trouvain2022-APV} define these types of pauses as ``gaps at turn changes in conversations'' and do not regard them as typical speech pauses  that are defined as ``pauses in connected speech section''. 
Regarding speech production and the temporal structure of speech, pauses also play a crucial role \citep{Trouvain2022-APV}. 
%
We have to distinguish between silent pauses and filled pauses such as ``uh'' or ``uhm'' \citep{Batliner95-FPI,Bilac2017-GAF,Trouvain2022-APV}. \citet{Bilac2017-GAF} extract \ac{MFCC} audio features and apply \acp{SVM} and \ac{RF} as classification methods.
Silent speech pauses and silence in audio can long since also be automatically detected, though \citep{Xu2020-LTS,Iqbal2018-GPA}.

Power, described by  \citet{Foxcabane13_TCM} as high competence due to skills, abilities,  or intelligence, can mainly be detected from audio by analysis of features related to fluency, such as speech rate and pauses. \citet{Luzardo14-EPS} perform an automated evaluation of student presentation skills and found a formant-based detection of filled pauses useful for classifying the overall quality of presentations. Further, they observed that speech rate is positively correlated with a speaker's self-confidence. A similar approach based on detecting filled pauses is taken by \citet{Ochoa18-RSA} in the audio modality of their automatic feedback system for presentation skills.

The mimicry of a conversation partner can help establish a connection in dyadic interactions. This may happen either subconsciously, or deliberately to project greater warmth and presence. \citet{Amiriparian2019_SII} investigate `synchronisation' (\ie the mutual adaptation of conversation partners) 
in such dyadic conversations,
 using acoustic and linguistic features on a dataset with $394$ speakers of six different cultures. For the acoustic analysis, both handcrafted \textsc{eGeMAPS} and deep \textsc{DeepSpectrum} features are extracted. Autoencoders are then used to measure the degree of synchronicity by training on one person and then reconstructing on their conversation partner. As the conversation continues, the reconstruction error tends to decrease across the six cultures, indicating that speakers are mutually adapting to each other.  

\subsubsection{Language}

A textual analysis lacks the information of prosody from a speech signal and must instead focus on linguistic cues. For purely text-based empathy and warmth recognition, we highlight two applications here: mental health support and stereotyping in social media. For presence, we also examine synchronicity in conversations.

\citet{Sharma2020} investigated empathy in the context of seeker-response interactions on text-based support platforms. Their framework, adapted for asynchronous communication, includes three mechanisms: emotional reactions to the seeker, interpretations conveying understanding, and explorations to improve understanding. A dataset of interactions collected from TalkLife and mental health subreddits was annotated in terms of empathy and rationales (text snippets motivating the empathy annotation). Then, a multi-task model based on two pre-trained \textsc{RoBERTa} encoders acting on seeker and response posts and a single attention layer combining their embeddings was proposed. The inclusion of seeker post and attention was found beneficial while fine-tuning the encoders and adding the rationale task gave minor improvements. 

Stereotypes are frequently encountered in social media posts, and may positively or negatively shape opinions on groups. \citet{Fraser2022_CMO} apply the warmth and competence model to stereotype identification by constructing a synthetic training set and building a model that can identify stereotypes in crowd-sourced data. First, using a seed lexicon, polar directions for warmth and competence are defined in a word embedding subspace. Then, sentences are created via templates filled with words of known polarity from the lexicon. For the word embeddings, \textsc{RoBERTa} is used, with \textsc{Glove} vectors serving as the baseline. 
A combination of \textsc{RoBERTa} embeddings with intermediate dimensionality reduction via partial Least-Squares performed best. Also, the generation of  sentences combining both warmth and competence-associated words improved accuracy by increasing the orthogonality of training pairs.

 \cite{Amiriparian2019_SII}
use \textsc{word2vec} to extract embeddings. The conversations are split into two parts, and the cosine similarity between their embeddings is computed. In addition, the co-occurrence of words between subjects in each part  is counted and normalised with the total number of words. The word embeddings showed little synchronisation (\ie mutual adaptation of conversation partners)   compared to the audio features, possibly indicating that the effect was happening too gradually on the linguistic level to measure during the short conversations. Word usage showed a clearer correlation but strongly differed across cultures, being most pronounced in British subjects.

\subsubsection{Other Modalities}

In order to recognise a charismatic person,  it is obvious to also consider other modalities, such as videos, images, and tactile sensors, as we mentioned earlier. 
In this context, videos and images might be especially beneficial for recognising how present and involved a person is in conversations by considering eye contact and facial expressions.

Some studies already try to use eye tracking in order to analyse attention and gaze patterns during social interactions~\citep{Rogers2018-UDE,Vehlen2021-EOA}. \citet{Rogers2018-UDE} also point out that some people show a preference towards mouth gaze, some for eye gaze, and others tend to vary the extent of their gaze between eyes and mouth. 
The authors apply a standard remote infrared eye tracker, consisting of an infrared sensor and a corresponding camera. \citet{Vehlen2021-EOA}, on the other hand,  employ special eye-tracking glasses enabling the opportunity for real-world experiments. In order to avoid expensive high-end processing devices, \citet{Zdarsky2021-ADL} introduce a \ac{CNN}  relying on video frames from low-cost web cameras. Another study also aims at making eye tracking available for everyone owning a mobile device with a camera \citep{Krafka2016-ETF}.

Besides our eyes, facial expressions are a very important tool to express excitement and emotions. \citet{Erez2008-STH} state that charismatic leaders exhibit more aroused behaviours than non-charismatic leaders.
For instance, smiles are arguably among the most visible and frequent markers and can convey a feeling of warmth and intimacy but also of fear or compliance~\citep{Awamleh2003-TAM}. There are already several approaches for automatic \ac{FER}, most of them utilising \ac{DL} and some sort of \ac{CNN} in particular~\citep{Li2020-DFE,Revina2021-ASO,Minaee2021-DEF,Wang2020-SUF}. 
The rough pipeline is to feed an input face image to a trained network and obtain a probability for a certain emotion category, such as happy or sad. In addition to using \acp{CNN} as the basic architecture blocks, there are extensions to improve performance, \eg adding an attention mechanism to the network~\citep{Li2018-OAF}.


\subsection{Automatic Generation of  Charisma}


We can approach the task of automatically generating charisma in two different ways. On the one hand, we can use an approach to try to imitate the charismatic characteristics of people previously defined in the literature. Charismatic persons are -- as outlined above -- characterised by a certain way of speaking (\eg pitch, duration, or rhythm during the conversation). The characteristics assigned to charismatic individuals can be obtained from previous works, such as \citet{Davidson21-TVO} or \citet{Klofstad16-HVP} and the many listed above, and thus represent an expert-based definition. Using this information in combination with generative machine learning methods, precisely these properties can be enforced when generating speech, resulting in a more charismatic perception.
On the other hand, methods such as reinforcement learning can also be used to generate charisma. The advantage of using this method is that new, previously unknown charismatic factors can be learnt.

\subsubsection{Expert-based Generation of Charisma}

To simulate charismatic behaviour, the previously identified building blocks must be taken into account when generating spoken language (or other modalities). In recent years, progress has been made in two main areas: First, generative methods \citep{Borsos22-AAL} for creating completely new audio outputs, and second, constrained audio generation, as well as style transfer, approaches, \eg \citep{Zhang19-LLR,Huzaifah20-AVD,Manzelli18-CDG}, where an existing audio file is stylistically adapted to pre-defined properties.

The latest results of generative methods for speech such as AudioLM are almost indistinguishable from real speech by humans \citep{Borsos22-AAL}. The high audio quality of the generated samples paves the path for further charismatic audio generation. Based on this, style control and  style transfer approaches can be used to change certain features of the voice \citep{Zhang19-LLR,Huzaifah20-AVD}. For example, \citet{Baird19-CDG} analysed if deep generative audio can be emotional. In doing so, they changed pitch as an important speech characteristic. In a similar way, other features can be adapted, leading to a more charismatic voice. 

In addition to the audio modality, this approach can be extended to other modalities, such as video or text. Based on findings from previous work investigating which features are perceived as particularly charismatic in the respective modality, these constraints can be considered in generative methods. For example, \citet{Ghorbani2022-ZZE} explore gesture generation from speech. Based on this work, charismatic gesture features can be taken into account, resulting in an overall charismatic perception of an AI such as by a virtual agent.

\subsubsection{Learning-based Generation of Charisma}

Automatically generating charisma can also be formulated as a weakly supervised machine learning task. For example, reinforcement learning methods have become increasingly popular in recent years in audio processing (and beyond) and are based on rewarding desired and punishing undesired behaviours \citep{latif22-ASO}. 

Applied to charisma generation, various characteristics of the speech are exploratively tried during generation. In doing so, the reward function includes usually indirect feedback from users on how charismatic the generated output is perceived. For example, pitch shifting, ranging from a low up to a very high pitch, can be explored. Taking user feedback into account, the optimal pitch that is perceived as most charismatic (or seems to be so, as it best solves a task that is best solved with high charisma) can be determined. 
In addition to direct user feedback, automatic charisma recognition approaches can be applied as a reward function to evaluate whether the generated behaviour is charismatic. In the context of generating emotional speech, \citet{Liu21-RLE} present such a paradigm. In a reinforcement learning setting, they train an automatic text-to-speech model to generate speech with emotions that can be discriminated by an automatic speech emotion recognition model. Another advantage of reinforcement learning is that new, so far unknown charismatic traits can be discovered using this method. This could range from obvious charismatic traits to entirely new charismatic behaviours that are as yet undiscovered in the literature and no one has thought of before, \eg as in \citet{Baker19-ETU} in a different context.

Obviosuly, in addition to the audio modality, reinforcement learning for charisma generation can be similarly applied to video and text and beyond.  
For instance, it might be beneficial for robots or virtual humans/agents to imitate charismatic gestures and appearance in general. In another use case, \citet{Won2021-CSF} have already  physically simulated humanoids performing competitive two-player sports, boxing and fencing, in a high degree-of-freedom environment. The applied control policies generated responsive and natural-looking behaviours \citep{Won2021-CSF}.

\section{Motivation and Applications of Computational Charisma}
\label{sec:applications}

A general motivation of charisma as a feature has been addressed in the introduction -- we now, however, turn to more specific examples of application use cases. 
%

As outlined above, charisma is often associated with charm and `magnetism', empowering to influence and inspire. Even if AI is not `experiencing' or showing charisma as humans could, it can be trained or programmed to appear charismatic. This leads to a plethora of use cases which motivate its realisation, but also often come with a number of risks and dangers. Below, we list examples broadly grouped. 

Let us first consider different aspects of communication: 
AI empowered with charisma may lead conversations with humans in potentially more convincing ways, engaging them, and potentially influence them towards decisions in favour of the AI's goals. Similarly, it can simulate empathy and in general be perceived as more intelligent due to socioemotional intelligence skills. Charimsatic techniques may improve everyday conversations by creating an emotional connection within communication partners or followers, make someone appear more powerful, competent, and worthy of respect \cite{Antonakis12_CCB}. This holds in conversations, but also in public speeches via means of AI in the future -- potentially to large audiences via the internet or in real-life settings. 
In automatic translation, AI could help translate charismatically or preserve charismatic traits in the target language. 
If AI is used for communication analysis, such as in mediation between human conversational partners, it can sense who is being more charismatic, more influencing, or more affected by the other party in a somewhat quantitative and subjective neutral manner.


Let us now turn to aspects of leadership: 
The great majority of use cases in research but also real-world applications is the training of charisma for enhanced leadership. More specifically, leaders in a variety of fields as politics, religion, or business highly benefit from being persuasive and likeable to achieve specific team goals and overcome potential resistance within employees or party members. Leaders may benefit greatly to win the trust of followers,  manage delicate operations, punish and reward, and achieve their goals \citep{Antonakis12_CCB}. \citet{Levay10_CLI} even argues that charismatic leaders are invariably proponents of change. Charisma may help AI inspire and motivate teams -- including encouraging and guidance through difficult challenges. Beyond, charisma can help AI to build and bind teams and communities in the first place. Especially in times of remote work and digital delivered social interaction as video calls and emails, such qualities of team coherence and trust and commitment are highly challenged. More specifically, team leaders may be fed back easily in their communication style with an AI that recognises charismatic language and enhances team outcomes, putting the team members at ease and comfort by simultaneously persuading them of a new idea. However, obtaining charisma -- including by humans that are trained on charisma skills by AI -- should not be abused for unethical goals (see section ethics of computational charisma below). Furthermore, a charismatic AI could also recruit new staff potentially more successfully than a non-charismatic one. However, it can also help it in negotiations within teams or with other parties in persuasive ways. If AI assists in decision making, charisma may help it to get the decisions across to the humans it presents it to.


Healthcare next appears suited field for charismatic AI: 
Be it for mental health or general health care -- charismatic AI could provide a compassionate, empathetic, and reassurance and support providing communication and assistance during diagnosis or interventions and therapy. 
Recognising charisma can also provide benefits for improving mental health. Mental health issues are widespread, affecting nearly one in five people worldwide \citep{Holmes2018_TLP} and incurring enormous human suffering and economic costs. The majority of patients prefer therapy to medication \citep{Holmes2018_TLP}, requiring the development of new solutions to improve the effectiveness of treatment. 
Depending on the application, those systems may either work in real-time for live session support or in an offline fashion for reviewing progress. A system that can process interactions and estimate the charismatic content may be a valuable tool for training professional practitioners. An effective therapist or counsellor can provide a sense of engagement and empathy to the patient, which in terms of charismatic dimensions involves both warmth (to show sympathy) and competence (to comprehend and engage with a patient's problems). In particular, this holds in case patients are reluctant or not sufficiently committed to overcome adverse feelings that sometime go along with behaviour change as for instance, confronting oneself with an anxiety-eliciting situation or the acquisition of new behaviours that feel initially uncomfortable. Both, therapists and clients could benefit from the rapport, empathy, believe of the ability to help, or persuasiveness of the therapist that go along with charisma. 
An AI solution that can help review and train these skills will likely also be beneficial for the increasingly popular digital support platforms, to improve the qualification of responders \citep{Sharma2020}. Although therapists are usually trained in providing ease, comfort, and being empathetic and receive supervision in doing so, the deployed speech in regard of influence and affability may remain neglected, although there is a great potential. Providing AI-generated automated feedback on the interaction and conversation style between therapist and client may improve the communication style and hence, therapy outcomes to a great extent. 
Considering the human-to-human conversation mediation alluded to above, this could include couple or other counseling endowed with active listening, and empathetic moderation.

Beside such fields, numerous other applications may benefit from charisma as an omnipotent quality:
%
%
In education, a charismatic AI may be more engaging and captivating for students taught by it. Furthermore, as a coach and mentor, charismatic AI could be more motivating. 
However, an AI that can sense and measure charisma can also help in tutoring about charisma, i.\,e., teach humans to be charismatic by monitoring their progress.
Further, in customer service, marketing and sales, charismatic AI could provide friendly and likable service, but also persuade customers.
%
In social media, charisma-empowered AI could interact with the social media users and generate a large group of followers, e.\,g., to influence opinions or provide positive brand connotation. This would include charismatic interaction with followers, responses to public reactions, or creation of new content in engaging ways.
 AI that has an understanding of charisma can also analyse charismatic behaviour and skills of users -- be it, e.\,g., for scientific analyses or identifying potential influencers early on.
%
As to gaming, non-player-characters (NPCs) driven by AI could be charismatic in oncoming games leading to an increasingly immersive gaming sensation. 
When it comes to event management and hospitality, 
charisma-empowered AI could provide engaging interaction with attendees of future virtual events. These may encompass a wide variety of events reaching from online conferences and workshops to virtual fair trades, virtual tours of galleries, museums, real-estate properties, or virtual concerts, festivals, parties, and other entertainment events. 
In particular, this could even include fundraising at suited events, where a charismatic AI could interact with donors persuasively. 
Beyond, hospitality at virtual or real-world occasions including checking guests at hotels in and out, question answering, recommendation giving, and furthermore, could be realised more charismatic by accordingly enabling future AIs.


More generally, any form of embodied or virtual AI -- e.\,g., assistive or companion agents -- could largely benefit from charismatic skills in the interaction with their users. This could help them in their communication and motivation as well as companionship including with lonely individuals.
 Across use cases, charismatic AI may be better positioned to personalise services to users by gaining access to their individual preferences, context, and history. 


\section{Ethics of Computational Charisma}
\label{ethics}

Of course, we do not aim at dark charisma for appealing to the baser human instinct; moreover, we do not want to harness `deceiving charisma',  \ie bright charisma for achieving goals that are per se unethical. 
As for spoken language as modality, a cover term might be `emotional speech’ in the sense of adding more credibility and user attachment to human-computer interaction. Charisma is thus not a goal in itself,  but a means to better achieve its goal.
Bright charismatic speech in itself cannot be unethical or ethical -- it always depends on the application we are envisioning. 
Thus, out of all the (ethical) cornerstones relevant for applications defined in \citep{Batliner22-EAI}, most might be  `secondary’ for charisma, \ie depend on the (type of) applications that use charismatic speech to pursue its goals. 
Yet, by providing charisma as a tool, we have to account for the possibility that this tool can be used for `dark goals' or, simply, that the outcome is not favourable. 
Thus, ethical requirements can be higher. 
In the same way, dealing with vulnerable groups of course puts higher requirements on ethics, as for instance, \textit{privacy} and \textit{avoiding harm} are concerned \citep{Batliner22-EAG}.

A self-learning system can adapt to templates and users the same way as the chatbot Tay learnt racist language from its users \citep{Wolf17-WWS}. This is a problem for every empathic virtual agent \citep{Pamungkas19-ECA}. Both dark charisma and bright charisma employed for dark goals can be created unwillingly or on purpose. In the first case, not only do algorithmic measures have to be taken, and in both cases, society has to define red lines against them.
\citet{Guerini05-TEP} reason about different capabilities of persuasive agents in case of ethical dilemmas (conflicting goals): (i) detect them and pass them on to a human; (ii) compute a possible conduct and pass this on to a human for final decision; (iii) make  own decisions. So far, we cannot envision artificial agents capable of doing this kind of reasoning in a reliable way. Thus, two principles should be followed: first, ethical dilemmas should be avoided by design, and if they are encountered, the decision has to be passed on to a human.

The most prominent specific ethical requirement might be \textit{disclosure of automation}  \citep{Mohammad22-ESF},  
belonging to the ethical cornerstones \textit{transparency} and \textit{accountability}: An according charismatic AI application has to make clear that the user is not interacting with some human being but with a computer. This must not be done in the small print but in a way that is really visible to the user. 
Transparency and accountability seem to be the primary cornerstones that impact \textit{autonomy}, \ie  provide the possibility for the user to be aware of the artificial charisma the application / the agent is  equipped with.   Then comes \textit{intrusive}: ethical requirements are higher, the more intrusive the application is. 
\citet{Montemayor22-IPO} claim that genuine empathy in healthcare is not possible for AI because it cannot be really emotional. Yet, a charismatic agent might at least act \textit{as-if}   but we have to make it clear towards the patients that the AI (robots, avatars) is artificial and does not have emotions or empathy itself, in order to prevent this erroneous and dangerous attribution that even can lead the user to fall in love with such a charismatic artificial agent.
The illusion of humanness can create the `uncanny valley' effect \citep{Mori12-TUV} when emotional/charismatic agents are close to human but still not close enough, by that irritating the human interaction partner. Both this uncanny valley and a too perfect humanness might be avoided by explicitly mentioning the artificial character of the agent, or by creating it in such a way that its non-human character is evident.
Note that some authors argue, under certain premises, in favour of anthropomorphous robots \citep{Darling17-WJA} or deception-capable robots 
\citep{Isaak17-WLO}. Although possible benefits might be evident, it is not clear at all how any dishonest use of such robots -- or, in our case, charismatic agents - could be prevented if not banned from the beginning.

\section{Conclusion and Outlook}

We discussed a `blueprint' for a charisma-savvy Artificial Intelligence -- able to analyse human charisma and generate charismatic behaviour itself. To this end, we introduced the origin and concept of charisma. We then
 discussed functional aspects of charsima by psychological models, mainly introducing two concepts based on the factors influence and affability as well as the theoretical concept of power, presence, and warmth as pillars of charisma. We argued that charismatic behaviour can be acquired and presented a brief summary of the literature on its acquisition. We then moved to formal aspects giving specific details on charisma as portrayed in spoken language. The choice to first focus on this modality was made, as today's AI largely interacts via spoken language with humans, and other modalities such as facial expression or body posture are yet to gain relevance. We further outlined computational aspects of modelling charisma. Here, we summarised the small body of literature on the automatic recognition and generation of charisma for audio, language, but also other modalities. As to the generation of charisma, we highlighted two avenues: First, based on the findings in the literature summarised up to that point in this article, one could design a charisma-empowered AI based on expert knowledge. Alternatively, weakly supervised machine learning could be exploited by either active learning methods questioning users about charisma skills of an AI or even by learning reinforced. In the latter, an AI would gain charismatic skills -- potentially even such unknown to date to humans -- that would help better accomplish its goal by interacting with users in real-world tasks -- ideally at scale. We then moved towards the plethora of potential use-cases of charisma-enabled AI, before introducing major ethical concepts to be considered at all times.

 Given the state-of-knowledge on charisma and the state-of-play in today's AI, it seems perfectly possible to endow AI with charisma skills. Currently, the literature on using machine learning for the recognition or generation of charisma or traits thereof largely focuses on the individual in isolation related to fields such as Affective Computing. However, disciplines such as Social Signal Processing moved also the consideration of the interplay between communicating parties into the foreground, which can be crucial for modelling charisma, \eg when it comes to mimicry. Further, audio, text, and video have so far been mostly considered, but touch, and more general haptics, have been touched upon as well. In the future, other modalities including smell, and further biological signals could be included. Further, the loop between recognition and generation of charismatic behaviour might be fully closed learning charismatic input/output of an AI `end-to-end'. 

 Overall, we envision a plethora of use-cases with great value of charsima-savvy AI. However, weakly-supervised  AI learning from large data may easily lead to new charismatic behaviours found by AI potentially reflecting back on human-to-human charismatic behaviour. Charismatic AI may also empower `dark' purposes or lead to negative effects such as AI influencing voters, e-shoppers, getting users addicted to or fall in love with the AI, and many more. As a community, we have to always contribute best to assure positive usage and the protection of users -- including in particular also from a technical end. A blueprint therefore might be considerably more challenging. Let us be best prepared for the rapid advent of charismatic AI.  


\section*{Conflict of Interest Statement}

The authors declare that the research was conducted in the absence of any commercial or financial relationships that could be construed as a potential conflict of interest.

\section*{Author Contributions}

All authors contributed equally to writing, revising, and editing the manuscript and approved the final version of the manuscript.

\section*{Funding}
This work is supported by the DFG's Reinhart Koselleck project No.\,442218748 (AUDI0NOMOUS).
%




\bibliographystyle{frontiersinSCNS_ENG_HUMS} 
\bibliography{paper}





\end{document}